\documentclass[aps,prc,twocolumn,nofootinbib,amsmath,showpacs,superscriptaddress,floatfix]{revtex4}
\usepackage{graphicx,epsfig,longtable}
\usepackage{mathptmx}
\usepackage[light,first]{draftcopy} % ,bottomafter
\usepackage{epstopdf}  
\usepackage[pdftex]{hyperref}
\usepackage{wasysym}
\usepackage{multirow}
\usepackage{eqnarray,amsmath}

\newcommand{\beqy}{\begin{eqnarray}}
\newcommand{\eeqy}{\end{eqnarray}}
\newcommand{\bmlet}{\begin{subequations}}
\newcommand{\emlet}{\end{subequations}}
\newcounter{saveeqn}

\def\gsimeq{\,\,\raise0.14em\hbox{$>$}\kern-0.76em\lower0.28em\hbox  
{$\sim$}\,\,}  
\def\lsimeq{\,\,\raise0.14em\hbox{$<$}\kern-0.76em\lower0.28em\hbox  
{$\sim$}\,\,}  

\begin{document}

\title{Quasicontinuum $\gamma$-decay of $^{91,92}$Zr: \\ benchmarking indirect ($n,\gamma$)
cross section measurements for the $s$-process}

\author{M.~Guttormsen}
\email{magne.guttormsen@fys.uio.no}
\affiliation{Department of Physics, University of Oslo, N-0316 Oslo, Norway}
\author{S. Goriely}
\affiliation{Institut d'Astronomie et d'Astrophysique, Universit\'{e} Libre de Bruxelles,
Campus de la Plaine, CP-226, 1050 Brussels, Belgium}
\author{A.~C.~Larsen}
\affiliation{Department of Physics, University of Oslo, N-0316 Oslo, Norway}
\author{A.~G{\"o}rgen}
\affiliation{Department of Physics, University of Oslo, N-0316 Oslo, Norway}
\author{T.~W.~Hagen}
\affiliation{Department of Physics, University of Oslo, N-0316 Oslo, Norway}
\author{T.~Renstr{\o}m}
\affiliation{Department of Physics, University of Oslo, N-0316 Oslo, Norway}
\author{S.~Siem}
\affiliation{Department of Physics, University of Oslo, N-0316 Oslo, Norway}
\author{N.~U.~H.~Syed}
\affiliation{Department of Physics, University of Oslo, N-0316 Oslo, Norway}
\author{G.~Tagliente}
\affiliation{Istituto Nazionale di Fisica Nucleare, Bari, Italy}
\author{H.~K.~Toft}
\affiliation{Department of Physics, University of Oslo, N-0316 Oslo, Norway}
\author{H.~Utsunomiya}
\affiliation{Department of Physics, Konan University, Okamoto 8-9-1, Higashinada, Kobe 658-8501, Japan}
\author{A.~V.~Voinov}
\affiliation{Department of Physics and Astronomy, Ohio University, Athens, Ohio 45701, USA}
\author{K.~Wikan}
\affiliation{Department of Physics, University of Oslo, N-0316 Oslo, Norway}
\date{\today}

\begin{abstract}
Nuclear level densities (NLDs) and $\gamma$-ray strength functions ($\gamma$SFs) have been
extracted from particle-$\gamma$ coincidences of the
$^{92}$Zr($p,p' \gamma$)$^{92}$Zr and $^{92}$Zr($p,d \gamma$)$^{91}$Zr reactions  
using the Oslo method. The new $^{91,92}$Zr $\gamma$SF data,
combined with photonuclear cross sections, cover the whole energy range from
$E_{\gamma} \approx 1.5$~MeV up to the giant dipole resonance at $E_{\gamma} \approx 17$~MeV.
The wide-range $\gamma$SF data display 
structures at $E_{\gamma} \approx 9.5$~MeV,
compatible with a superposition of
the spin-flip $M1$ resonance and a pygmy $E1$ resonance. Furthermore,
the $\gamma$SF shows a minimum at $E_{\gamma} \approx 2-3$~MeV and an
increase at lower $\gamma$-ray energies.
The experimentally constrained NLDs and $\gamma$SFs are
shown to reproduce known ($n, \gamma$) and Maxwellian-averaged cross sections for
$^{91,92}$Zr using the {\sf TALYS} reaction code, thus serving as a benchmark for 
this indirect method of estimating ($n, \gamma$) cross sections for Zr isotopes.
\end{abstract}

%\pacs{21.10.Ma, 27.50.+e, 25.40.Hs}
% PACS Numbers: 21.10.Ma (level density), 
%               21.10.-k (Properties of nuclei; nuclear energy levels), 
%               21.60.Jz (HF & QRPA)
% 		25.20.Lj (Photoproduction reactions)
%		24.30.Gd Other resonances
%		25.55.Hp	3He transfer reactions
%		27.40.+z	39 ≤ A ≤ 58
%		27.50.+e	59 ≤ A ≤ 89
%		25.40.Hs	Transfer reactions, nucleon-induced
% From Voinov's PRL: 25.40.Lw, 25.20.Lj, 25.55.Hp, 27.40.+z

\maketitle

\section{Introduction}
\label{sec:int}
The interplay of the microscopic, quantum-mechanical regime and
the macroscopic world is crucial for many physical systems.
In nuclear astrophysics, various stellar environments and extreme
cosmic events represent the playground for the nucleosynthesis,
for which nuclear properties determine the outcome together with the
astrophysical conditions.

For elements heavier than iron, two neutron capture processes~\cite{BBFH57,Cameron57} 
dominate their creation. These two processes are characterized by the timescale, 
rapid ($r$) and slow ($s$),
in comparison with the $\beta^{-}$-decay rates.
Typically, the neutron energies are in the $0.01-1$ MeV range, corresponding to stellar
temperatures of $0.1-10$ GK. 
The $r$-process, although the astrophysical site is not
yet firmly established~\cite{Arnould2007}, takes place at such high
neutron densities ($> 10^{20}$ cm$^{-3}$)
that the neutron capture process totally dominates the competing $\beta^{-}$ decay
until the neutron flux is exhausted.
The $s$-process ($T \sim 0.1$ GK and neutron density $\approx10^8$ cm$^{-3}$)
operates at much longer time scales allowing
for $\beta^{-}$ decay prior to the next neutron capture~\cite{BBFH57,Cameron57,Arnould2007}.

The weak $s$-process is believed to take place in massive stars ($M>8M_{\odot}$~\cite{kappeler2011})
and produces most of the $s$-abundances in the mass region between Fe and Zr,
while the main $s$-process operates in Asymptotic
Giant Branch (AGB) stars and produces the heavier $s$-process isotopes up to the lead/bismuth 
region~\cite{kappeler2011}. 

The neutron capture cross section is small for isotopes with
magic neutron numbers. This results in bottlenecks for the reaction flow,
giving rise to the buildup of sharp abundance maxima. This is reflected in the
solar-system abundances: we find $s$-process peaks at mass numbers $A \approx 90, 140$ and $210$
corresponding to the magic neutron numbers $N = 50, 82$ and $126$, respectively~\cite{Arnould2007}.
%The $s$- and $r$-processes at the $A \approx 90$ mass region are particularly important for
%producing seeds to build heavier elements~\cite{Raiteri1993,Gallino1998,Arlandini1999,Lugaro2003}.
% CHECK REFERENCES!

A crucial question is whether the nuclear system after neutron absorption will keep
the neutron and emit $\gamma$ rays to dissipate the energy, or rather eject the neutron or
other particles/fragments and thereby produce other elements. For the $s$-process,
this may happen at the so-called \textit{branch points}, where the $\beta^-$-decay rate 
is comparable with the $(n,\gamma)$ rate. 
The relative probability to keep the neutron depends strongly on the nucleus'
ability to emit $\gamma$ rays, which is governed by the $\gamma$-ray strength function ($\gamma$SF)
and the nuclear level density (NLD) of the compound system.

For zirconium isotopes, with semimagic proton number $Z=40$, and at/close to the $N=50$ closed shell, 
neutron-capture cross sections are typically low and one could question whether statistical approaches
such as the Hauser-Feshbach framework~\cite{hauser1952} is applicable. Further, this is the meeting point
of the weak and main $s$-process, and although $^{96}$Zr traditionally has been considered an 
$r$-process isotope, it could be significantly produced~\cite{Lugaro2003} through neutron 
capture on the branch point nucleus $^{95}$Zr, depending on its $(n,\gamma)$ cross section 
and the neutron flux at the $s$-process site.

However, since $^{95}$Zr is unstable with a half-life of 64 days, 
no direct $(n,\gamma)$ cross-section measurement has been
performed to date, and so only theoretical estimates are available. 
Recent work has
discovered unexpected enhancements in the $\gamma$SF of several zirconium isotopes, 
such as the $E1$ pygmy dipole resonance as well as strong $M1$ transitions
close to neutron threshold~\cite{utsunomiya2008,Iwamoto2012}. 
The presence of such enhanced $\gamma$-decay
probabilities in $^{95}$Zr could boost its neutron-capture rate.

Several applications may take advantage of better knowledge of the
NLDs and $\gamma$SFs in the $A \approx 90$ mass region.
The production and destruction rates of $^{93}$Zr is interesting for the interpretation
of the relative abundance of the radioactive $^{93}$Nb and $^{93}$Zr pair,
which can be used to estimate the $s$-process temperature and also, 
together with the $^{99}$Tc-$^{99}$Ru pair, 
act as a chronometer to determine the time elapsed since the start of 
the $s$-process~\cite{Neyskens2015}. The $^{93}$Zr$(n,\gamma)$ cross section
has been measured up to 8 keV~\cite{tagliente2013}, but contributions to the 
Maxwellian-averaged cross section at higher energies are based on theoretical
calculations and would benefit from experimental constraints~\cite{lugaro2014}.

In this work, we report on the NLDs and $\gamma$SFs for the $^{91,92}$Zr isotopes
with neutron number $N=51$ and 52, respectively, and use our data as input for calculating the
$^{90,91}$Zr$(n,\gamma)$ cross sections with the reaction code {\sf TALYS}~\cite{Koning12}. 
As there exist direct $(n,\gamma)$ measurements
for these isotopes, we use these cases as a benchmark for
our indirect method of determining the $(n,\gamma)$ cross section
in this mass region. 
These investigations are
part of a larger campaign to study the branch-point 
neutron capture rates at the $A \approx 90$ $s$-process peak.

The outline of the present manuscript is as follows.
In Sect.~II the experiment and results are described.
The NLDs and $\gamma$SFs are extracted by means of the Oslo method
and compared with model calculations
in Sects.~III and IV, respectively.
In Sect.~V  radiative neutron capture cross sections using the {\sf TALYS} code and
experimental NLDs and $\gamma$SFs as inputs are compared with known cross sections.
A summary and an outlook are given in Sect.~VI.

\section{Experimental results}
\label{sec:exp}
The experiments were performed at the Oslo Cyclotron Laboratory (OCL) 
with 17-MeV and 28-MeV proton beams for
the $^{92}$Zr($p,p'$)$^{92}$Zr and $^{92}$Zr($p,d$)$^{91}$Zr reactions, respectively.
The target was a 2 mg/cm$^2$ thick metallic foil enriched to 95\% in $^{92}$Zr.

%---------------------------------------------------%
\begin{figure}[h]
\begin{center}
\includegraphics[clip,width=\columnwidth]{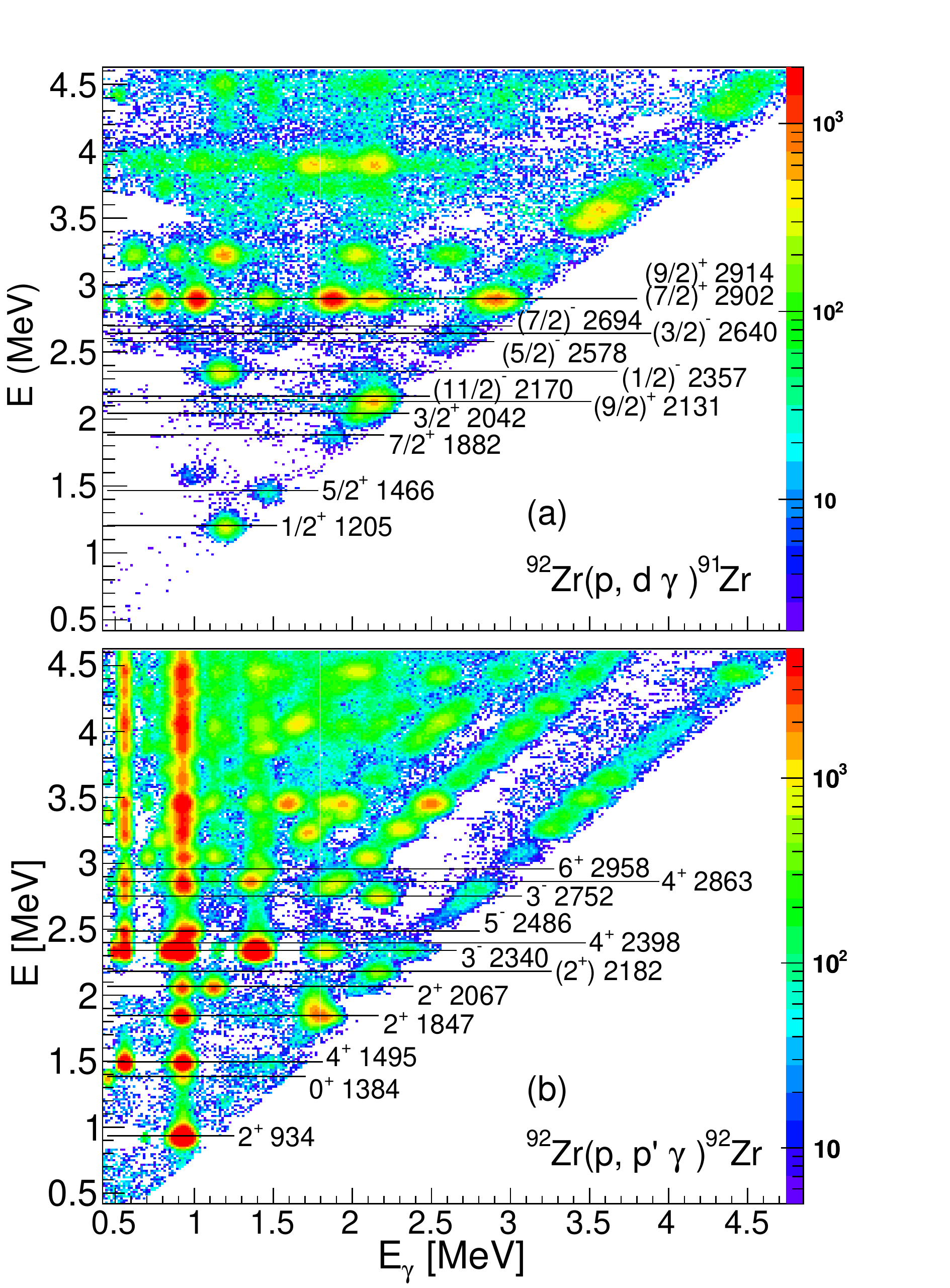}
%\vskip 2cm
\caption{(Color online) The discrete part of the particle-$\gamma$ coincidence matrices for the
(a) $^{92}$Zr($p,p'$)$^{92}$Zr and (b) $^{92}$Zr($p,d$)$^{91}$Zr reactions.
The pixel width is 16 keV $\times$ 16 keV.}
\label{fig:matrices}
\end{center}
\end{figure}
%---------------------------------------------------%

%---------------------------------------------------%
\begin{figure}[h]
\begin{center}
\includegraphics[clip,width=0.8\columnwidth]{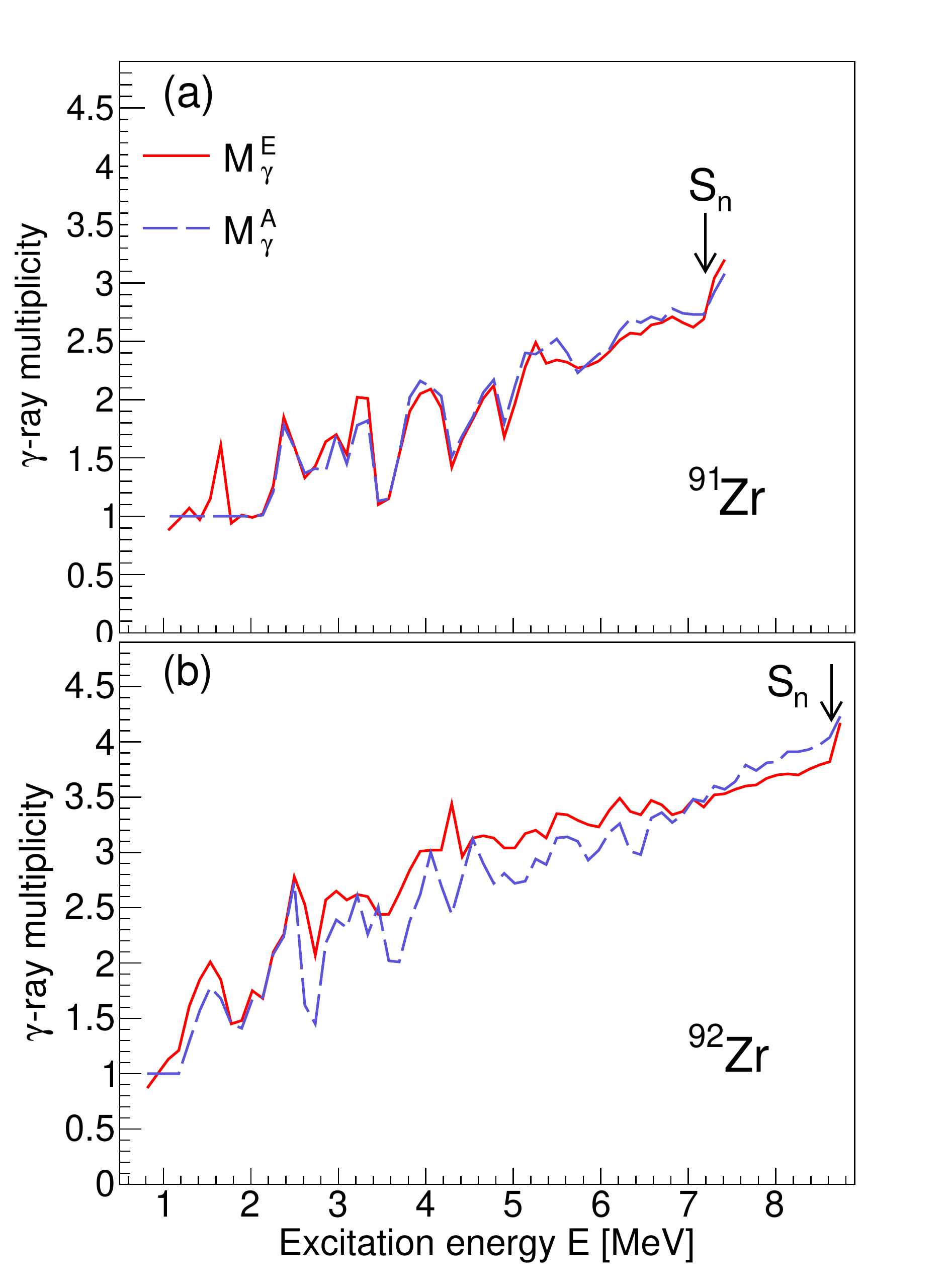}
%\vskip 2cm
\caption{(Color online) The $\gamma$-ray multiplicity measured as function of
excitation energy for the
$^{92}$Zr($p,p'$)$^{92}$Zr and $^{92}$Zr($p,d$)$^{91}$Zr reactions.
The energy bin is 120 keV.}
\label{fig:mult9192}
\end{center}
\end{figure}
%---------------------------------------------------%
The charged outgoing particles were measured with the SiRi system
of 64 $\Delta E - E$ silicon telescopes with thicknesses
of 130 and 1550 $\mu$m, respectively~\cite{siri}.
The Si detectors were placed in forward direction
covering $\theta = 42^\circ$ to $54^\circ$
relative to the beam. The typical energy resolutions
measured with the telescopes were 75 and 95 keV full-width half maximum for the
($p,p'$)$^{92}$Zr and ($p,d$)$^{91}$Zr reactions, respectively.
By setting 2-dimensional gates on the two ($E, \Delta E)$ matrices,
the outgoing charged ejectiles for the desired reactions were selected.
Coincident $\gamma$ rays for the residual $^{91,92}$Zr were measured with the CACTUS
array \cite{CACTUS} consisting of 28 collimated $5" \times 5"$ NaI(Tl)
detectors with a total efficiency of $14.1$\% at $E_\gamma = 1.33$ MeV.

%---------------------------------------------------%
\begin{figure*}[t]
\begin{center}
\includegraphics[clip,width=1.8\columnwidth]{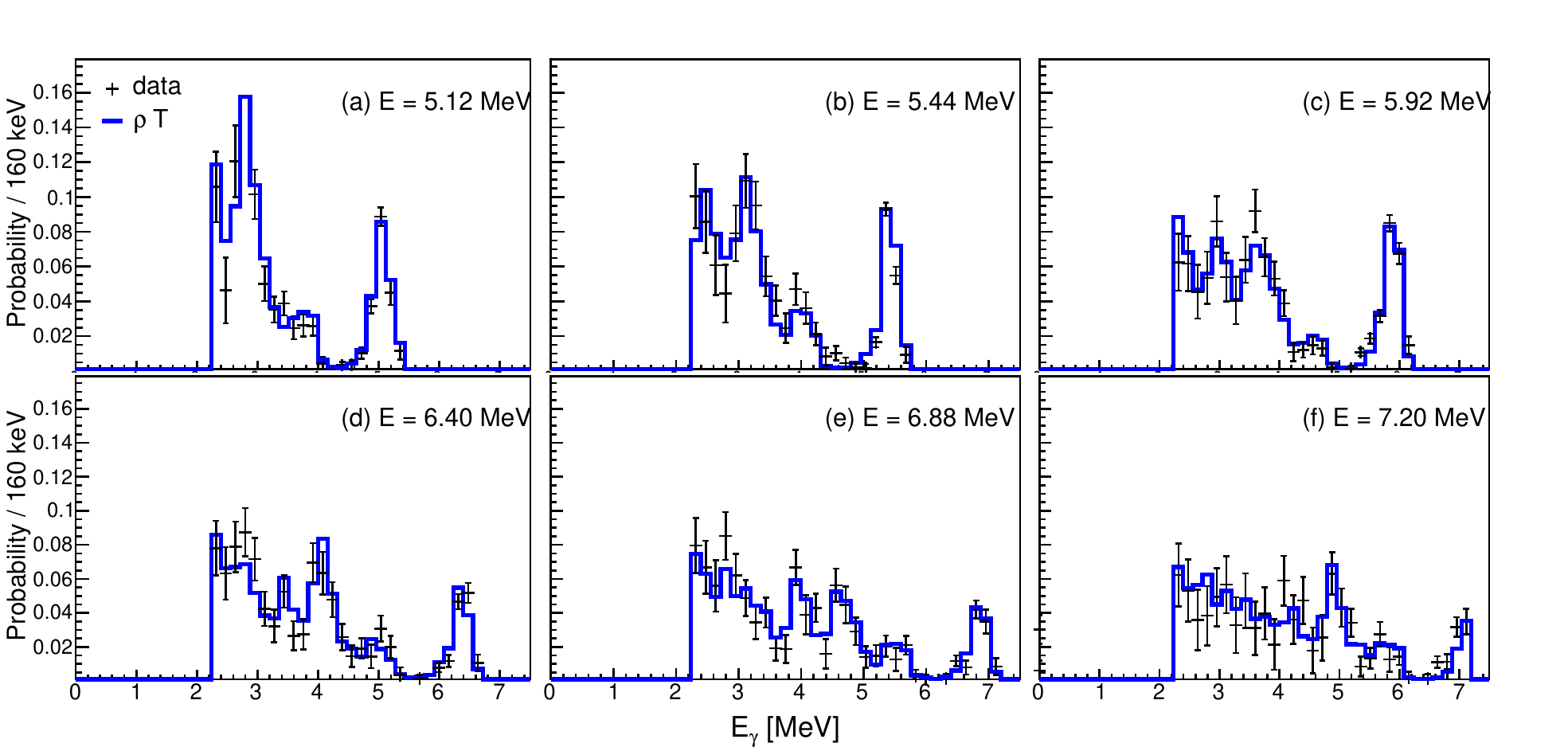}
%\vskip 2cm
\caption{(Color online) Primary $\gamma$-ray spectra from various
initial excitation energies $E$ (crosses) in $^{91}$Zr. The spectra
are compared to the product
$\rho(E-E_{\gamma}) {\mathcal{T}}(E_{\gamma})$ (blue histogram).
Both the $\gamma$ and excitation energy dispersions are 160 keV/ch.}
\label{fig:does91}
\end{center}
\end{figure*}
%---------------------------------------------------%
%---------------------------------------------------%
\begin{figure*}[t]
\begin{center}
\includegraphics[clip,width=1.8\columnwidth]{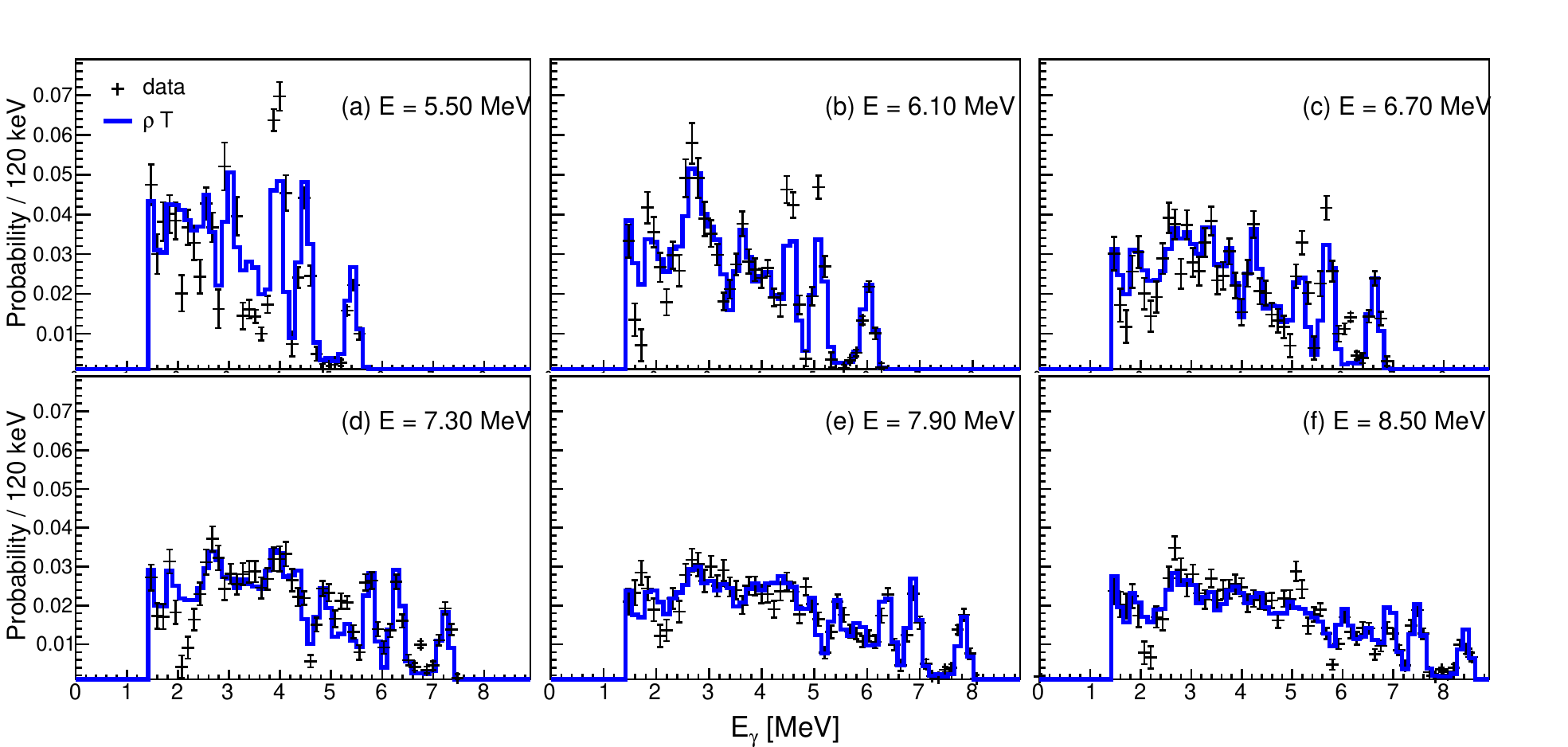}
%\vskip 2cm
\caption{(Color online) Same as Fig.~\ref{fig:does91} for $^{92}$Zr.
Both the $\gamma$ and excitation energy dispersions are 120 keV/ch.}
\label{fig:does92}
\end{center}
\end{figure*}
%---------------------------------------------------%

The first step in the analysis is to sort the $\gamma$-ray spectra as function of excitation energy.
Knowing the details of the reaction kinematics,
the excitation energy $E$ is given by the energy of the outgoing charged particle.
Figure~\ref{fig:matrices} shows the discrete part of the
particle-$\gamma$ matrices $(E_{\gamma}, E)$ for the residual $^{91,92}$Zr
with prompt coincidence requirements. The $\gamma$-ray spectra for each
excitation energy has been unfolded with new NaI-response functions.
The unfolding procedure~\cite{Gut96} has proven to work very well also
for continuum $\gamma$-ray spectra.

The $\gamma$-ray multiplicity as function of initial excitation energy $E$
can be calculated by the energy ($E$) and area ($A$) methods~\cite{Gut87}:
\begin{eqnarray}
M^{E}_{\gamma}(E) &=& \frac{E}{\langle E_{\gamma}(E) \rangle},
\label{eq:energy}\\
M^{A}_{\gamma}(E) &=& \frac{A_{\rm total}(E)}{A_{\rm primary}(E)},
\label{eq:area}
\end{eqnarray}
where
$\langle E_{\gamma}\rangle$ is the average energy of the total $\gamma$-ray spectrum,
$A_{\rm total}$ is the intensity (area) of that spectrum, and
$A_{\rm primary}$ is the intensity of the primary $\gamma$-ray spectrum.

Figure ~\ref{fig:mult9192} shows the $\gamma$-ray multiplicities
$M^{E}_{\gamma}$ and $M^{A}_{\gamma}$ for
the two reactions. There are practically no statistical
errors in the data points due to the high number of counts in the $\gamma$-ray spectra.
Since $M^{E}_{\gamma}$ represents the most direct and transparent method,
we use this quantity as the measure for the multiplicity, and
$M^{A}_{\gamma}$ as an indicator for systematical errors. We find
significant deviation between the two methods only at a few excitation energies.
For example a 60\% deviation is found at $E \approx 1.6$~MeV in $^{91}$Zr, which is due
to a weak contamination peak of $E_{\gamma}\approx 1$~MeV
located where no levels are expected for $^{91}$Zr.

The discrete part of the $\gamma$-ray matrices in Fig.~\ref{fig:matrices}
are characterized by
isolated peaks in the two-dimensional landscape spanned by the initial excitation energy $E$
and the $\gamma$-ray energy $E_{\gamma}$. Several peaks
tend to fall onto diagonals in the matrices. The diagonal with
$E_{\gamma} \approx E$ represents decay directly to the ground state with
a $\gamma$-ray multiplicity of $M_{\gamma} = 1$. We also recognize
vertical and horisontal lines in the matrices. The vertical lines
correspond to yrast transitions from the last steps in the $\gamma$-ray cascades.
The horizontal lines appear when levels have high $\gamma$-ray multiplicity or
several levels are bunched together in excitation energy.

For excitations below $E \approx 3$ MeV, most of the levels and
$\gamma$ transitions are easily recognized by comparing
with known data from literature~\cite{NNDC}. For $^{91}$Zr, we
see that levels with spin/parity $I^{\pi}$ from $1/2^{+}$ up to ($9/2^{+}$) or even ($11/2^{-}$)
are populated in the ($p,d$) reaction, as also reported by Blok {\em et al.}~\cite{blok76}.
The population of the latter two high-spin states are due to $\ell = 4$ and $(5)$
transfer, probably involving the $g_{9/2}$ and $h_{11/2}$ neutron orbitals.
A peculiar situation is seen for the (1/2)$^{-}$ 2357 keV level, which shows
only one peak at $E_{\gamma} \approx 1.18$ MeV. The peak is actually the composition
of two transitions with almost the same $\gamma$ energies (1152 keV and 1205 keV).
This is consistent with the $\gamma$-ray multiplicity of $M_{\gamma} \approx 2$
shown in Fig.~\ref{fig:mult9192}.

Figure~\ref{fig:matrices} (b) shows that the inelastic proton reaction on $^{92}$Zr
populates a broad spin window ranging from $0^+$ to $6^+$.
Since the ground state spins of $^{91,92}$Zr are $5/2^+$ and $0^+$, Fig.~\ref{fig:mult9192}
reveals about one unit more of multiplicity for $^{92}$Zr compared to $^{91}$Zr. Only levels of the lowest
part of the spin distribution of $^{92}$Zr can directly decay to the $0^+$ ground state.
This is manifested by the dominant feeding into the
diagonals of the first excited $2^+$ and $4^+$ states in Fig.~\ref{fig:matrices} .
We also observe the vertical lines corresponding to
the transitions $4^+\rightarrow 2^+$ (561 keV) and $2^+ \rightarrow 0^+$ (934 keV).
As an example, the multiplicity spectrum of $^{92}$Zr has
a peak with $M_{\gamma} \approx 2$ at $E\approx 1.5$ MeV. This is mainly due to
the first $4^+$ state that decay via the $2^+$ state into the ground state,
giving $M_{\gamma} = 2$. At about the same excitation energy,
Fig.~\ref{fig:matrices} (b) shows the decay path
of the first $0^+$ state, which also goes via the $2^+$ state
giving multiplicity $M_{\gamma} = 2$.

The energy distribution of first-generation or
primary $\gamma$ rays can be extracted from the unfolded total $\gamma$-ray spectra
of Figs.~\ref{fig:matrices} (a) and (b). Let $U^E(E_{\gamma})$ be the
unfolded $\gamma$-ray spectrum at a certain initial excitation energy $E$.
Then the primary spectrum can be obtained by a subtraction of a weighted sum of
$U^{E'}(E_{\gamma})$ spectra for $E'$ below $E$:
\begin{equation}
F^E(E_{\gamma})=U^E(E_{\gamma}) - \sum_{E' < E}w_{E'}U^{E'}(E_{\gamma}).
\end{equation}
The weighting coefficients $w_{E'}$ are determined
by iterations as described in Ref.~\cite{Gut87}.
After a few iterations, the multiplicity of the primary spectrum should be $M(F^E) \approx 1$, where the
multiplicity of the total spectrum is determined by $M(U^E)=M^{E}_{\gamma}(E)$ from Eq.~(\ref{eq:energy}).
The obtained primary spectra are organized into a matrix $P(E_{\gamma}, E)$
that is normalized according to $\sum_{E_{\gamma}}P(E_{\gamma}, E)=1$.

The next step of the Oslo method, is the factorization
\begin{equation}
P(E_{\gamma}, E) \propto   \rho(E-E_{\gamma}){\cal{T}}(E_{\gamma}) ,\
\label{eqn:rhoT}
\end{equation}
where we assume that the decay probability is proportional to the
NLD at the final energy $\rho(E-E_{\gamma})$ according
to Fermi's golden rule~\cite{dirac,fermi}. The decay is also proportional
to the $\gamma$-ray transmission coefficient ${\cal{T}}$, which is assumed
to be independent of excitation energy according to the Brink hypothesis~\cite{brink,guttormsen2016}.

The relation (\ref{eqn:rhoT}) makes it
possible to simultaneously extract the two one-dimensional vectors $\rho$
and ${\cal{T}}$ from the two-dimensional landscape $P$. We use the
iteration procedure of Schiller {\em et al.}~\cite{Schiller00}
to determine $\rho$ and ${\cal{T}}$ by a least $\chi ^2$ fit
using relation (\ref{eqn:rhoT}). For this extraction, we have chosen the following part of the
$P$ matrix:
For $^{91}$Zr the excitation energy region is
5.0~MeV $< E < $ 7.2~MeV with $E_{\gamma}> 2.4$~MeV, and for $^{92}$Zr we choose
4.5~MeV $< E < $ 8.6~MeV with $E_{\gamma}> 1.5$~MeV.

The applicability of relation (\ref{eqn:rhoT}) and the quality of the fitting
procedure are demonstrated in Figs.~\ref{fig:does91} and~\ref{fig:does92} for $^{91,92}$Zr, respectively.
The agreement is satisfactory when one
keeps in mind that the $\gamma$-decay pattern fluctuates from level to level.
With the rather narrow excitation energy bins of 160 and 120 keV for $^{91,92}$Zr, respectively,
each $\gamma$-ray spectrum will be subject to significant
Porter-Thomas fluctuations~\cite{PT} responsible for
local deviations for individual primary spectra
compared to the global average given by $\rho{\cal{T}}$.
It should be mentioned that only the spectra from a few excitation
energy bins are shown, however, all spectra show the same agreement with $\rho{\cal{T}}$.
Further tests and justification of the Oslo method have been discussed in Ref.~\cite{Lars11}.

\section{The nuclear level density}
\label{sect:nld}

The functional form of $\rho$ and $\cal{T}$ are uniquely identified through the fit, 
but the scale and slope of these functions are still undetermined.
It is shown in Ref.~\cite{Schiller00} that functions generated by the transformations:
\begin{eqnarray}
\tilde{\rho}(E-E_\gamma)&=&A\exp[\alpha(E-E_\gamma)]\,\rho(E-E_\gamma),
\label{eq:array1}\\
\tilde{{\mathcal{T}}}(E_\gamma)&=&B\exp(\alpha E_\gamma){\mathcal{T}} (E_\gamma)
\label{eq:array2}
\end{eqnarray}
give identical fits to the primary $\gamma$-ray spectra, as the examples shown in Figs.~\ref{fig:does91} and~\ref{fig:does92}.
In the following, we will estimate the parameters $A$ and $\alpha$ from systematics and other experimental data. 
The normalization of ${\mathcal{T}}$ by the
constant $B$, only concerns the $\gamma$SF that will be discussed in the next subsection.

%--------------------------------------------------------------------------%
%\begin{table}[]
%\caption{Experimental level spacings and average $\gamma$ widths
%for $\ell = 0$ and $1$ neutron capture experiments~\cite{Tagliente2008a,Tagliente2008b,RIPL3}.}
%\begin{tabular}{c|c|cc|cc}
%\hline
%\hline
%Nucleus   &$S_n$&      $D_0$ &      $D_1$ &$\langle\Gamma_{\gamma 0}\rangle$&$\langle\Gamma_{\gamma 1}\rangle$\\
%&(MeV)&        (eV)&        (eV)&               (meV)             &               (meV)             \\
%\hline
%$^{91}$Zr &7.195&   7179(233)&    2241(58)&                          181(14)$^*$)&                199(16)$^*$)          \\
%$^{92}$Zr &8.635&     514(15)&     260(20)&                          140(40)&                220(30)          \\
%\hline
%\hline
%\end{tabular}
%\\
%\label{tab:parameters1}
%$^*$) Unrealistic error bars, see text.
%\end{table}
%--------------------------------------------------------------------------%

%--------------------------------------------------------------------------%
\begin{table*}[]
\caption{Experimental level spacings and average $\gamma$ widths
for $\ell = 0$  neutron capture experiments~\cite{Tagliente2008a,Tagliente2008b}.
For the evaluation of $\langle\Gamma_{\gamma 0}\rangle$ of $^{91}$Zr,
additional resonances from Ref.~\cite{MugAtlas} were added, see text.}
\begin{tabular}{lcccccc}
\hline
\hline
Nucleus &$S_n$ [MeV]&$D_0$ [eV]&$\langle\Gamma_{\gamma 0}\rangle$ [meV]      & $\langle\Gamma_{\gamma 0}\rangle$ [meV]                          &$\langle\Gamma_{\gamma 0}\rangle_B$ [meV]      & $\langle\Gamma_{\gamma 0}\rangle_B$ [meV]\\
Ref.    &           &\cite{Tagliente2008a,Tagliente2008b}&\cite{MugAtlas}    & \cite{Capote09}                                                       &\cite{Tagliente2008a,Tagliente2008b,MugAtlas}& adopted                                \\

\hline
$^{91}$Zr &7.195&   7179(233)&  170(20)  & 130(20) &            180(137) & 130(40)    \\
$^{92}$Zr &8.635&     514(15)&  140(40)  & 134(16) &          131(56)    & 140(40)    \\
\hline
\hline
\end{tabular}
\\
\label{tab:parameters1}
\end{table*}
%--------------------------------------------------------------------------%

The normalization of the NLD is determined by known levels at low excitation energies
and the NLD at the neutron separation energy $\rho(S_n)$ which can be estimated from the $s$-wave
resonance spacing $D_0$ \cite{Tagliente2008a,Tagliente2008b},
as listed in Table~\ref{tab:parameters1}.
However, such an extraction requires knowledge of the spin and parity distributions
of the NLD at the neutron separation energy and is consequently model-dependent.
For these reasons, two different NLD formulations are considered,
namely the constant-temperature (CT) formula \cite{Koning08,Capote09}
and the Hartree-Fock-Bogolyubov (HFB) plus combinatorial model \cite{Goriely08} (with
ptable and ctable {\sf TALYS} parameters), which give two quite different descriptions of the
energy, spin and parity dependences of the NLD.
In the case of the HFB plus combinatorial model, the NLD is tabulated and its spin and parity
distributions determined by the underlying effective interaction. The total NLD $\rho(S_n)$ deduced
from the $s$-wave resonance spacings are given in Table~\ref{tab:parameters3}.
Note that the HFB plus combinatorial model predicts that the NLD equiparity is achieved
only above the neutron separation energy, at typically 9~MeV, in both $^{91,92}$Zr.

%--------------------------------------------------------------------------%
\begin{table}[]
\caption{Total $^{91,92}$Zr NLDs at the neutron separation energy deduced with the HFB plus combinatorial
model from the experimental s-wave spacing $D_0$ and its uncertainties indicated by low (L), recommended (R) and high (H).}
\begin{tabular}{lccc}
\hline
\hline
Nucleus  & $\rho_{\rm L}(S_n)$&$\rho_{\rm R}(S_n)$&$\rho_{\rm H}(S_n)$  \\
         & [MeV$^{-1}$]&     [MeV$^{-1}$] &      [MeV$^{-1}$]  \\
\hline
$^{91}$Zr& 7200        &   7440       &    7700        \\
$^{92}$Zr&19560        &  20120      &  20700   \\
\hline
\hline
\end{tabular}
\label{tab:parameters3}
\end{table}
%--------------------------------------------------------------------------%

In contrast, the CT formula  is bound to assume an equiparity distribution and to follow a spin distribution given by ~\cite{Ericson}
\begin{equation}
g(E,I) \simeq \frac{2I+1}{2\sigma^2(E)}\exp\left[-(I+1/2)^2/2\sigma^2(E)\right],
\label{eq:spindist}
\end{equation}
where $E=S_n$ at the neutron binding energy, $I$ is the spin and
$\sigma(E)$ the energy-dependent spin cutoff parameter, which turns out to be
the main contributor to the uncertainties in the estimate of the total NLD.
The spin cutoff parameter $\sigma$ is traditionally determined by a close-to rigid moment of inertia.
Since $\sigma^2= \Theta T/\hbar^2$~\cite{Ericson} and the nuclear temperature $T$
is assumed to be approximately constant for $2\Delta < E < S_n$~\cite{luciano2014,CT2015}, $\sigma^2$
follows the energy dependence of the moment of inertia $\Theta$. We assume that $\Theta$ is proportional to
the number of quasiparticles, which again is proportional to $E$. Thus, we
write
\begin{equation}
\sigma^2(E)=\sigma_d^2 + \frac{E-E_d}{S_n-E_d}\left[\sigma^2(S_n)-\sigma_d^2\right],
\label{eq:sigE}
\end{equation}
which goes through  two anchor points. The first point $\sigma_d^2$
is determined from known discrete levels at excitation energy $E=E_d$.
The second point at $E=S_n$ is estimated assuming a rigid moment of inertia~\cite{egidy2005}:
\begin{equation}
\sigma^2(S_n) =  0.0146 A^{5/3} \cdot \frac{1+\sqrt{1+4aU_n}}{2a},
\label{eqn:eb}
\end{equation}
where $A$ is the mass number, and $U_n=S_n-E_1$ is the intrinsic excitation energy.
The level NLD parameter $a$ and the energy shift parameter $E_1$ is determined according to Ref.~\cite{egidy2005}.

In order to obtain a systematic error band, we multiply
the rigid moment of inertia $\Theta_{\rm rigid}=0.0146 A^{5/3}$ of Eq.~(\ref{eqn:eb})
with a factor $\eta$, which takes the values $\eta =0.6$, 0.8 and 1.0 for the
low (L), recommended (R) and high (H) values, respectively. The
corresponding spin cutoff parameters
and NLDs are listed in Table~\ref{tab:parameters2}.

Comparing Tables~\ref{tab:parameters3} and \ref{tab:parameters2},
the HFB plus combinatorial model predicts significantly higher total NLD
at $S_n$ that can hardly be taken into account by the parameter uncertainties in
the CT approach. Both approaches will consequently be considered in the present analysis,
not only for determining the NLD, but also the corresponding $\gamma$-ray strength function, as detailed below.

%--------------------------------------------------------------------------%
\begin{table*}[]
\caption{Parameters used to extract NLDs within the CT model approach. Systematical uncertainties
are indicated by low (L), recommended (R) and high (H) values (see text).}
\begin{tabular}{c|cc|cc|ccc|ccc|cc}
\hline
\hline
Nucleus  & a           & $E_1$&$E_d$&$\sigma_d$&$\sigma_{\rm L}(S_n)$&$\sigma_{\rm R}(S_n)$&$\sigma_{\rm H}(S_n)$&$\rho_{\rm L}(S_n)$&$\rho_{\rm R}(S_n)$&$\rho_{\rm H}(S_n)$ &$T_{\rm CT}$ &   $E_0$ \\
         &[MeV$^{-1}$] & [MeV]&[MeV]&      &         &                     &                     & [MeV$^{-1}$]&      [MeV$^{-1}$] &     [MeV$^{-1}$]  &    [MeV]&   [MeV] \\
\hline
$^{91}$Zr& 9.84       &-0.03& 2.5& 3.1(2)  &3.83 &  4.42               & 4.95               & 4230(140)        &    5590(180)       &    6950(230)        &  0.88(5)& -0.29(48)\\
$^{92}$Zr&10.44       & 0.66& 3.0& 3.0(2)  &3.89 &  4.50               & 5.03               &13500(390)        &   16640(490)       &   19840(580)        &  0.90(2)& -0.02(21)\\

%$^{91}$Zr& 9.835       &-0.033& 2.5& 3.1(2)  &3.830 &  4.423               & 4.945               & 4228(137)        &    5591(182)       &    6953(226)        &  0.88(5)& -0.29(48)\\
%$^{92}$Zr&10.442       & 0.659& 3.0& 3.0(2)  &3.894 &  4.496               & 5.027               &13500(394)        &   16640(486)       &   19840(579)        &  0.90(2)& -0.02(21)\\

\hline
\hline
\end{tabular}
\label{tab:parameters2}
\end{table*}
%--------------------------------------------------------------------------%

%---------------------------------------------------%
\begin{figure}
\begin{center}
\includegraphics[clip,width=\columnwidth]{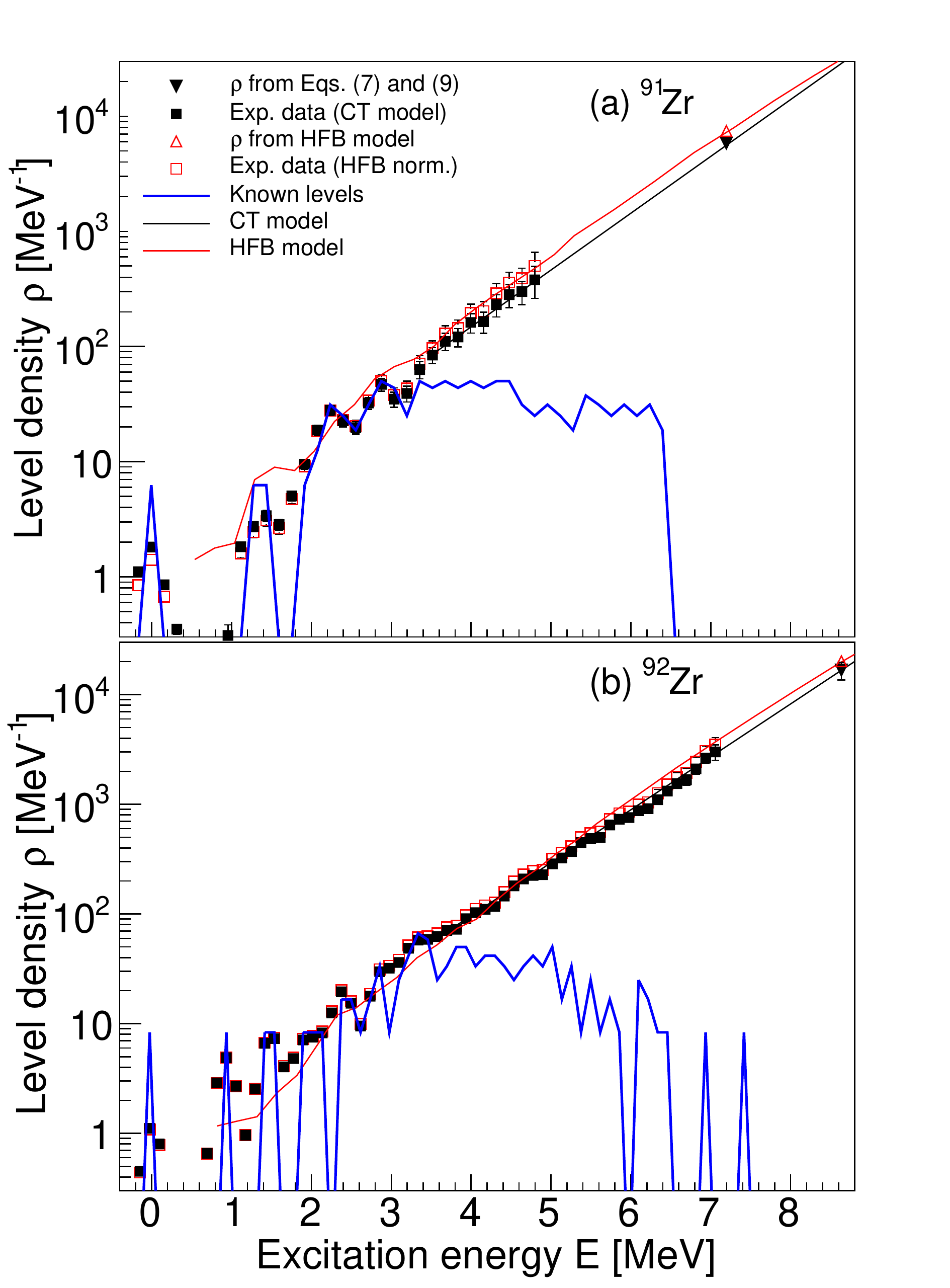}
%\vskip -1cm
\caption{(Color online) Normalized NLDs of $^{91,92}$Zr extracted from the present experiment.
At low excitation energies, the data are normalized
to known discrete levels (solid blue line). At higher excitation energies,
the data are normalized to $\rho$ at $S_n$ using resonance spacing $D_0$.
Two normalization procedures for  $\rho(S_n)$ are used: (i) the spin
cutoff parameter listed in Table~\ref{tab:parameters2} together with the
CT NLD parameters (black line and symbols) and
(ii) the HFB plus combinatorial model (red line and symbols)~\cite{Goriely08}.}
\label{fig:counting_9192}
\end{center}
\end{figure}
%---------------------------------------------------%

%---------------------------------------------------%
%\begin{figure}
%\begin{center}
%\vskip -2cm
%\includegraphics[clip,width=\columnwidth]{fig_nld_hfb_91Zr.pdf}
%\vskip -6.1cm
%\includegraphics[clip,width=\columnwidth]{fig_nld_hfb_92Zr.pdf}
%\vskip -2cm
%\caption{(Color online) Same as Fig.~\ref{fig:counting_9192} for $^{91}$Zr where the
%CT NLD formula is replaced by the HFB plus combinatorial (solid red line) \cite{Goriely08} model.}
%\label{fig:counting_9192b}
%\end{center}
%\end{figure}
%---------------------------------------------------%

When $\rho(S_n)$ is estimated, we still need to bridge the
energy gap between our data points and the estimated $\rho(S_n)$ value. To do so,
we use the corresponding NLD formula, {\it i.e.} the HFB plus combinatorial model in
the first case and the CT formula in the second case~\cite{Ericson}
\begin{equation}
\rho(E)=\frac{1}{T_{\rm CT}}\exp\left({\frac{E-E_0}{T_{\rm CT}}}\right) ,
\label{eq:ct}
\end{equation}
where the temperature $T_{CT}$ and energy shift $E_0$ are free parameters
adjusted to the data and given in Table~\ref{tab:parameters2} for the two Zr isotopes.

The experimental NLDs for $^{91,92}$Zr are shown in Fig.~\ref{fig:counting_9192}
for the CT approach and the HFB plus combinatorial model. In both cases,
a rather CT pattern is found for the total NLD above typically 3~MeV,
though their respective slopes are different following different predictions of the total NLD at $S_n$.

\section{The $\gamma$-ray strength function}
\label{sect:gsf}

The standard way to determine the remaining normalization coefficient $B$ of Eq.~(\ref{eq:array2}) is
to constrain the data to the known total radiative
width $\langle \Gamma_{\gamma 0} \rangle$ at $S_n$ \cite{Schiller00,voin1}, defined as 
\begin{eqnarray}
\langle\Gamma_{\gamma 0} (S_n)\rangle=\frac{1}{2\pi\rho(S_n, I, \pi)} \sum_{I_f}&&\int_0^{S_n}{\mathrm{d}}E_{\gamma} {\cal {T}}(E_{\gamma})
\nonumber\\
&&\times \rho(S_n-E_{\gamma}, I_f),
\label{eq:norm}
\end{eqnarray}
where the summation and integration run over all final levels with spin $I_f$ that are accessible by $E1$ or $M1$
transitions with energy $E_{\gamma}$. This procedure is known to work well
when the individual $\gamma$ widths are centered around a common average value.

%--------------------------------------------------------------------------%
\begin{figure}[]
\begin{center}
%\vskip -2cm
\includegraphics[clip,width=\columnwidth]{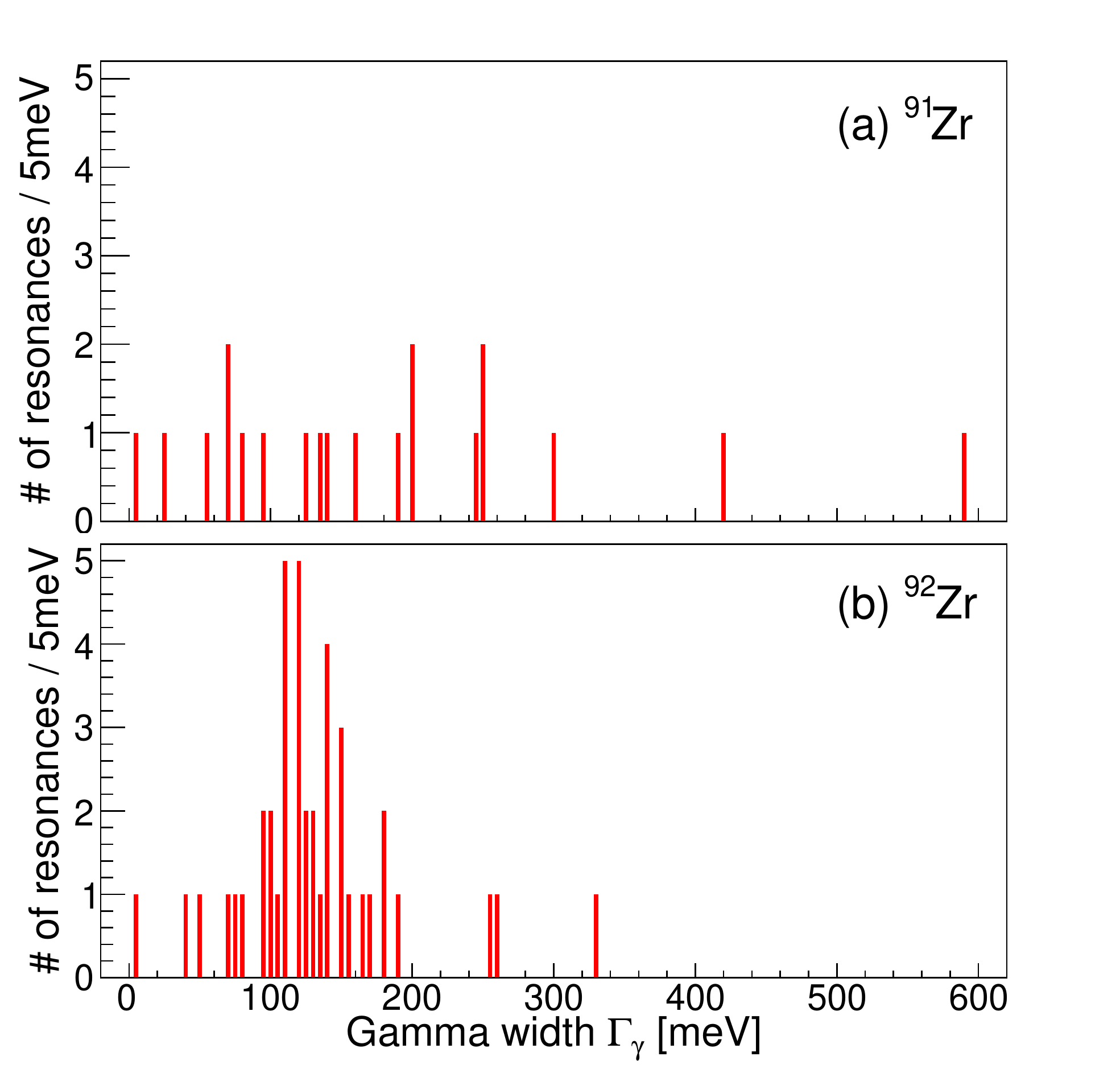}
%\vskip -2cm
\caption{(Color online) The number of $\ell=0$ resonances as function of their specific $\gamma$ width $\Gamma_{\gamma}$.
For $^{91}$Zr, 15 resonances are taken from Ref.~\cite{Tagliente2008a} and additional 5 from Ref.~\cite{MugAtlas}.
For $^{92}$Zr, the 42 resonances are taken from Ref.~\cite{Tagliente2008b}. The data are used to
re-evaluate the average $\gamma$ width, called $\langle\Gamma_{\gamma 0}\rangle_B$ in column 6 of
Table~\ref{tab:parameters1} .
}
\label{fig:gg_9192}
\end{center}
\end{figure}
%--------------------------------------------------------------------------%

%--------------------------------------------------------------------------%
\begin{figure}[]
\begin{center}
\includegraphics[clip,width=\columnwidth]{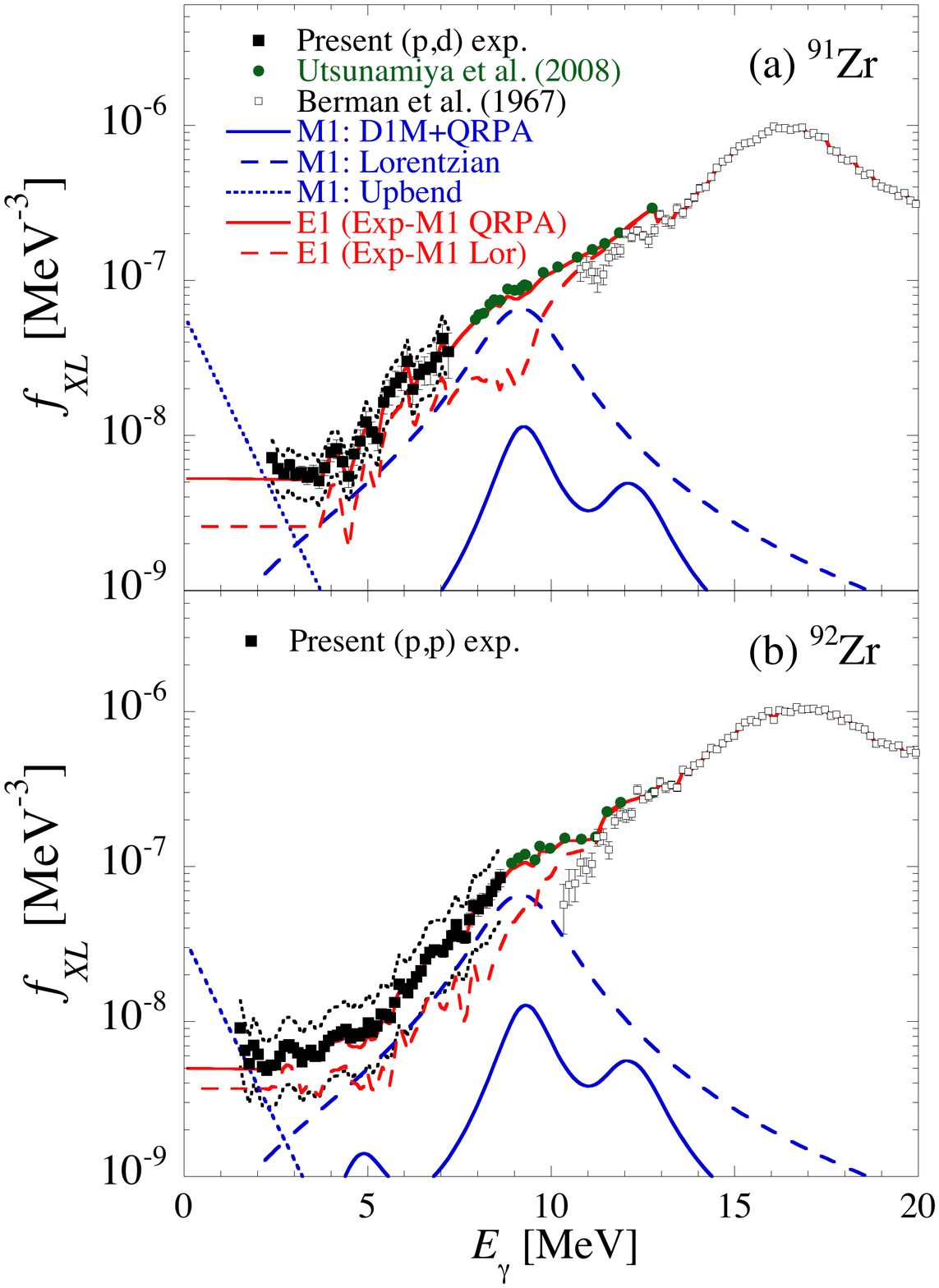}
%\includegraphics[scale=0.4]{fig_gsf_Zr.pdf}
%\vskip 2cm
\caption{(Color online) (a) Experimental $^{91}$Zr $\gamma$SFs (solid squares) with its error band (dotted lines)
due to the uncertainty in $\sigma$, $D_0$ and $\langle\Gamma_{\gamma 0}\rangle$
(see Table~\ref{tab:parameters2}). Also shown are the photoneutron data \cite{utsunomiya2008,berman1967},
the spin-flip $M1$ strength derived from either  D1M+QRPA calculations (solid blue line) or
a Lorentzian function (dashed blue line) and the $M1$ upbend (dotted blue line).
The resulting recommended $E1$ strength is shown by a solid (dashed) red line
and obtained from the total experimental dipole strength by subtracting the
D1M+QRPA (Lorentzian) $M1$ component. (b) Same for the $^{92}$Zr $\gamma$SFs.}
\label{fig:gsf_Zr}
\end{center}
\end{figure}
%--------------------------------------------------------------------------%

Columns 4 and 5 in Table~\ref{tab:parameters1} list the experimental $\langle\Gamma_{\gamma 0}\rangle$
values from literature~\cite{MugAtlas,Capote09}. However, Fig.~\ref{fig:gg_9192} shows that
the individual $\gamma$ widths scatter much more than the experimental uncertainties for the
individual $\gamma$ widths, which are usually below $\approx 20$ meV. For $^{91}$Zr,
we can hardly locate a common average $\langle\Gamma_{\gamma 0}\rangle$ as the data scatter from 5.5 to 590 meV.
Thus, the standard method of calculating weighted average and uncertainties
give unrealistic small errors in the case of $^{91}$Zr.

Therefore, we calculate instead the mean average of the $\gamma$ widths
and the standard deviation of these values by
\begin{eqnarray}
\langle\Gamma_{\gamma 0}\rangle_B       &=& \frac{1}{n}\sum_{i=1}^{n} {\Gamma_{\gamma 0}^i},
\label{eq:average}\\
\Delta\langle\Gamma_{\gamma 0}\rangle_B &=& \sqrt{\frac{1}{n-1} \sum_{i=1}^{n} ({\Gamma_{\gamma 0}^i} - \langle\Gamma_{\gamma 0}\rangle_B )^2},
\label{eq:std}
\end{eqnarray}
where $n$ is the number of resonances.
The index $B$ indicates that these values are relevant for the determination of the coefficient $B$.
For $^{92}$Zr, we find reasonably
consistent values of the average $\gamma$ widths in columns 4, 5 and 6 of Table~\ref{tab:parameters1},
and adopt the value $\langle\Gamma_{\gamma 0}\rangle_B= 140\pm 40$~meV.
As expected, the uncertainty in the average $\gamma$ width for $^{91}$Zr is very large.

To constrain the $^{91}$Zr data further, we use the photonuclear reaction data \cite{utsunomiya2008,berman1967} around $S_n$
to determine the $B$ value.
The transformation from photonuclear cross section $\sigma_\gamma$ to $\gamma$SF is performed by~\cite{Capote09}:
\begin{equation}
f (E_{\gamma}) =\frac{1}{ 3\pi ^2 \hbar^2c^2}\frac{ \sigma_\gamma(E_{\gamma})} { E_{\gamma}}.
\end{equation}
Note that the photoneutron cross section in the direct vicinity of the neutron threshold is not
considered to estimate the corresponding $\gamma$SF, since in this region it
remains also sensitive to the neutron channel and the $\gamma$SF can consequently
not be deduced from the cross section in an unambiguous way.

In turn, the dipole $\gamma$SF, including both the $E1$ and $M1$ contributions,
can be calculated from our measured transmission
coefficient \cite{Capote09} through
\begin{equation}
f (E_{\gamma}) =\frac{1}{2 \pi}\frac{{\cal {T}}(E_{\gamma})}{ E_{\gamma}^3 }.
\end{equation}
The corresponding experimental $\gamma$SFs for $^{91,92}$Zr are displayed as solid squares in
Fig.~\ref{fig:gsf_Zr}. The figure also includes the $\gamma$SFs
derived from $^{91,92}$Zr($\gamma$, n) cross section data by
Utsunomiya {\em et al.}~\cite{utsunomiya2008} and Berman {\em et al.}~\cite{berman1967}.
As mentioned above, we have normalized our $^{91}$Zr data points to match the $(\gamma, n)$ data
at $S_n$, as shown in Fig.~\ref{fig:gsf_Zr}. The adopted $\langle\Gamma_{\gamma 0}\rangle_B$ values
used to normalize the $\gamma$SF and estimate the uncertainies are given in column 7 of Table~\ref{tab:parameters1}.

Since our dipole strength includes both the $E1$ and $M1$ contributions, for
estimating the average radiative width $\langle\Gamma_{\gamma 0}\rangle$ as well as the
radiative neutron capture cross section of $^{90}$Zr, they need to be disentangled,
especially in view of the non-equiparity of the NLD predicted within the HFB plus
combinatorial approach \cite{Goriely08}. For this purpose, we have estimated the
spin-flip $M1$ resonance from two different approaches, namely the HFB plus
Quasi-Particle Random Phase Approximation (QRPA) based on
the Gogny D1M interaction \cite{Goriely16} and a Lorentzian function,
both guided by a previous experimental analysis of photoneutron
measurements \cite{utsunomiya2008} as well as ($p,p'$) scattering
data on $^{90}$Zr close to $\theta=0$ degrees~\cite{Iwamoto2012}.
Such experiments revealed an $M1$ resonance located at a centroid
energy $E_{\rm M1} \simeq 9-9.5$~MeV with a width $\Gamma_{\rm M1} \simeq 2.50$ MeV.
At almost the same energies, an $E1$ pygmy resonance with $E_{\rm PDR1}=9.2$ MeV
and $\Gamma_{\rm PDR1} = 2.9$ MeV has been found. Such structures at around 9.5 MeV
have been reported also for the $^{92,94,96}$Zr isotopes~\cite{Tamii2015}.
For our sensitivity analysis, we consider both options, {\it i.e.} possible $M1$
representations, including a strong $M1$ Lorentzian with a
peak cross section $\sigma_0=7$~mb \cite{utsunomiya2008}
as well as the D1M+QRPA strength, as shown in Fig.~\ref{fig:gsf_Zr}. The D1M+QRPA strength is seen to
be significantly less than the phenomenological Lorentzian strength inferred in
Ref.~\cite{utsunomiya2008} giving rise to a stronger possible $E1$ counterpart.

Finally, our measurements at the lowest energies ({\it i.e.} around 2~MeV) also suggest the
presence of a low-energy enhancement (the so-called upbend) that has been suggested by shell model
calculations to be of $M1$ nature \cite{Brown14,schwengner2013}. 
For nuclei studied in this mass region with the Oslo method, we find a
low-energy enhancement (upbend) of the
$\gamma$SF~\cite{Guttormsen2005,Spyrou2014,Larsen2016,Renstrom2016,Tveten2016}. The upbend has
also been verified for $^{96}$Mo using another technique~\cite{Wiedeking2012}.

To describe the low-energy enhancement,
it is therefore important to include below 2~MeV an $M1$ upbend that may influence not only
the estimate of the total radiative width  $\langle\Gamma_{\gamma 0}\rangle$, but also the
radiative neutron capture cross section. The upbend structure is described by the exponential
function~\cite{schwengner2013,spyrou2014}
\begin{equation}
f_{\rm upbend}(E_{\gamma}) = C \exp (-\eta E_{\gamma}).
\end{equation}
The adopted parameters $C=5 \pm 2 \times 10^{-8}~{\rm MeV}^{-3}$ and $\eta=1.1 \pm 0.5$~MeV for
modelling the upbend of $^{91}$Zr and  $C=3.5 \pm 0.5 \times 10^{-8}~{\rm MeV}^{-3}$ and $\eta=1.1 \pm 0.5$~MeV  for $^{92}$Zr.

With this procedure, it is possible to disentangle from experimental data, the $E1$ and $M1$
components together with their relative model uncertainties for a sensitivity analysis.
The resulting $E1$ strengths deduced from the experimental strength by subtracting the  D1M+QRPA
or Lorentzian spin-flip $M1$ contribution as well as the low-energy $M1$ upbend, are shown
in Fig.~\ref{fig:gsf_Zr} for both Zr isotopes. Note that a constant $E1$ strength function
is assumed for energies $E_\gamma \rightarrow 0$, as indicated by shell model
calculations \cite{Sieja17} and empirically described by the generalized Lorentzian approach \cite{Kopecky90}.

With these resulting $E1$ and $M1$ strengths and NLD (as detailed in Sect.~\ref{sect:nld}),
we obtain for $^{91}$Zr a $\langle\Gamma_{\gamma 0}\rangle_B = 130 \pm 40$~meV, after
normalizing our $^{91}$Zr data points to match the $(\gamma, n)$ data at $S_n$. As mention above for  $^{92}$Zr,
we adopt $\langle\Gamma_{\gamma 0}\rangle_B = 140 \pm 40$ meV to constrain the experimental dipole strength.

\section{The radiative neutron capture cross sections}
\label{sect:xs}

%--------------------------------------------------------------------------%
\begin{figure}[]
\begin{center}
\includegraphics[clip,width=\columnwidth]{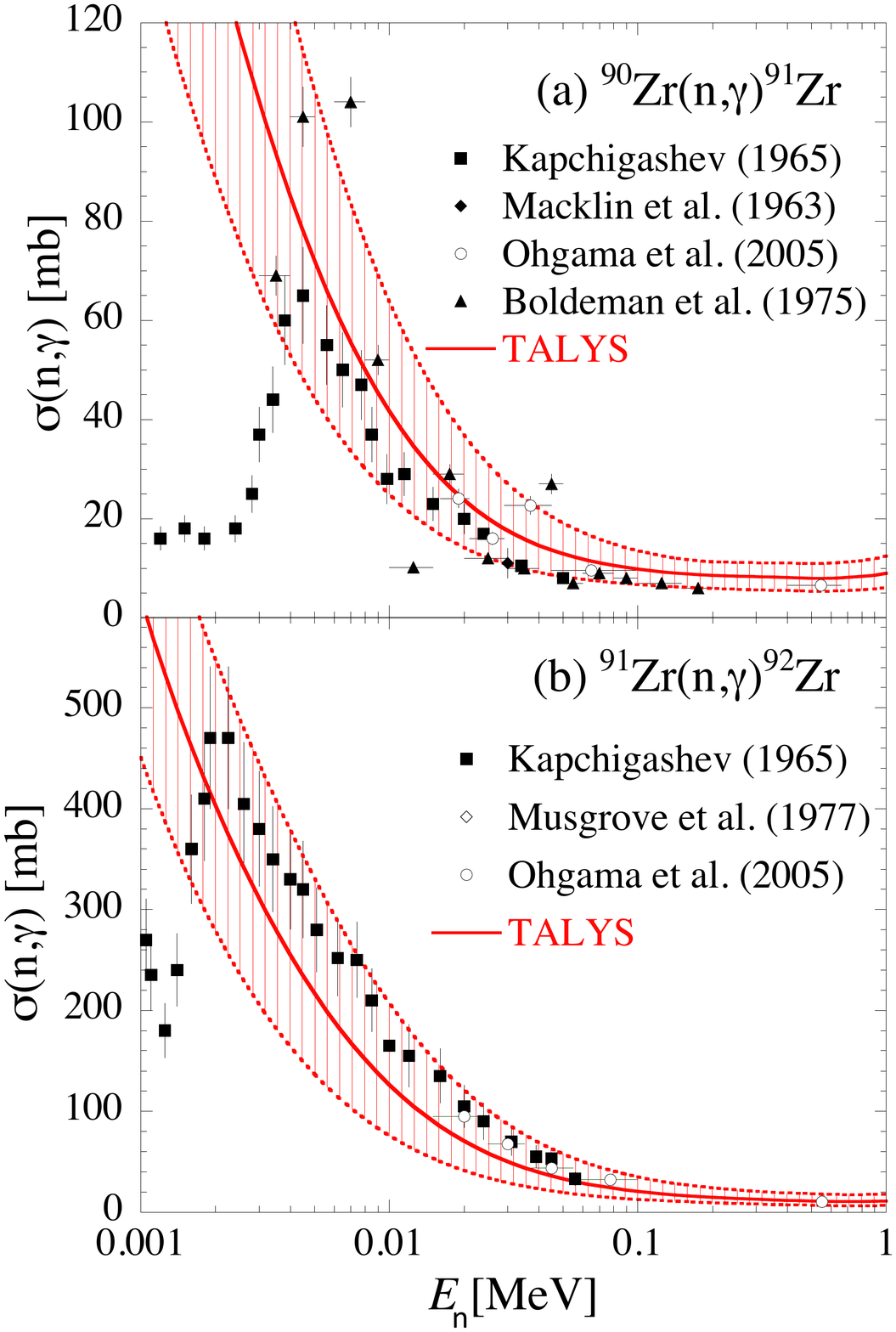}
%\vskip 2cm
\caption{(Color online) (a) Comparison between the experimental $^{90}$Zr($n,\gamma$)$^{91}$Zr
cross section~\cite{Kapchigashev65,Macklin63,Ohgama05} and the one obtained with the {\sf TALYS} code
on the basis of the NLD and $\gamma$SF derived experimentally in the present work.
The hashed area depicts all the experimental and
model-dependent uncertainties taken into account in the present analysis.
(b) Same for $^{91}$Zr($n,\gamma$)$^{92}$Zr cross section with experimental
data taken from Refs.~\cite{Kapchigashev65,Ohgama05,Musgrove77}. }
\label{fig:xs_Zr}
\end{center}
\end{figure}
%--------------------------------------------------------------------------%

%--------------------------------------------------------------------------%
\begin{figure}[]
\begin{center}
\includegraphics[clip,width=\columnwidth]{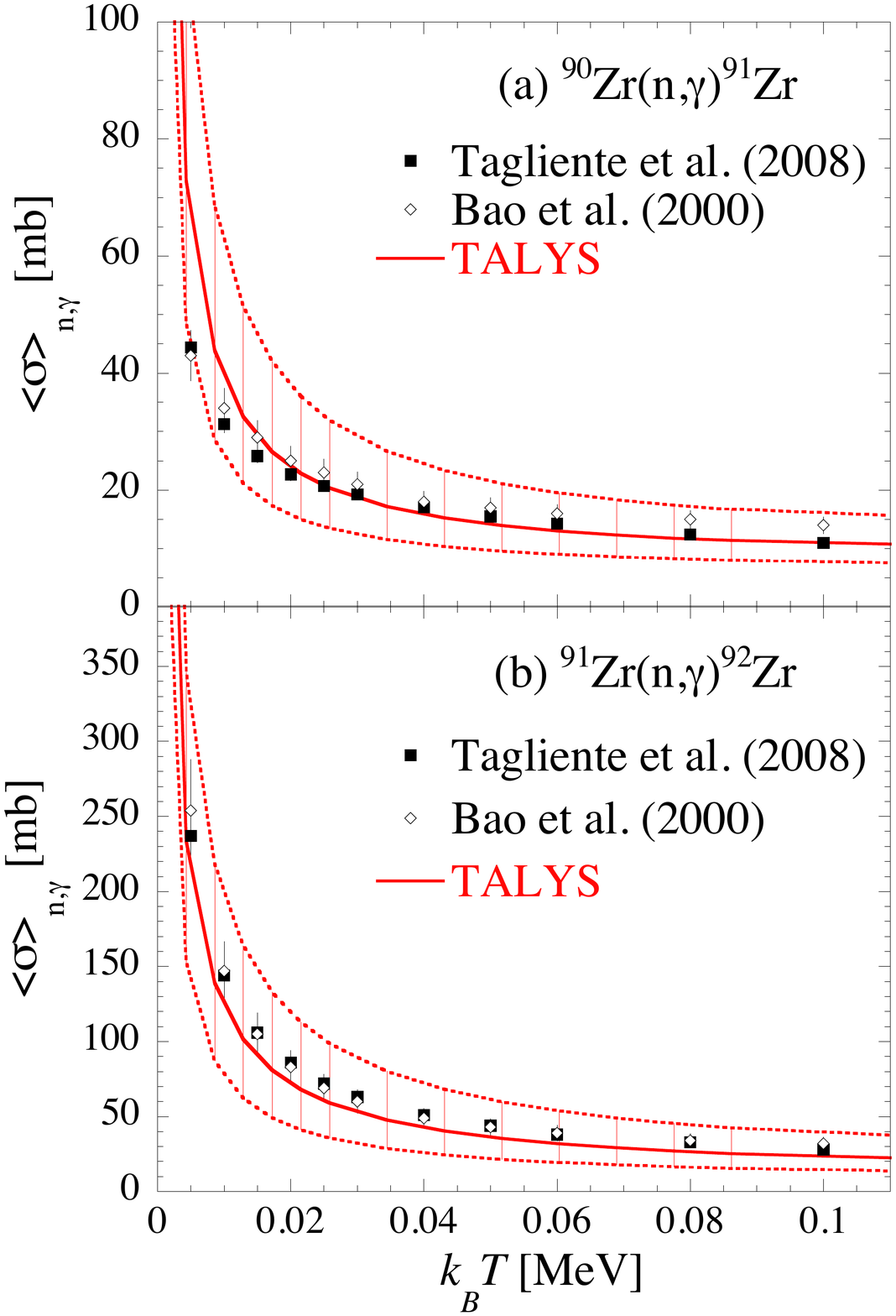}
%\vskip 2cm
\caption{(Color online) Comparison between the experimental $^{90}$Zr($n,\gamma$)$^{91}$Zr
Maxwellian-averaged cross section  \cite{Tagliente2008a,Bao00} and the one obtained
with the {\sf TALYS} code on the basis of the NLD and $\gamma$SF derived experimentally in the present work.
(b) Same for $^{91}$Zr($n,\gamma$)$^{92}$Zr Maxwellian-averaged cross section with
experimental data taken from Refs.~\cite{Tagliente2008b,Bao00}.}
\label{fig:macs_Zr}
\end{center}
\end{figure}
%--------------------------------------------------------------------------%

The NLD and $\gamma$SF derived in the previous sections can now be tested on the additional experimental
data relative to the radiative neutron capture cross sections $^{90}$Zr($n,\gamma$)$^{91}$Zr
and $^{91}$Zr($n,\gamma$)$^{92}$Zr.
These cross sections essentially depend on the photon transmission
coefficient of the final compound nucleus, hence to the NLD and $\gamma$SF obtained from
the present experiments. We compare in Fig.~\ref{fig:xs_Zr} the experimentally
known ($n,\gamma$) cross sections with the  theoretical calculations obtained
with the {\sf TALYS} reaction code \cite{Koning12}. Both cross section calculations use
directly the $E1$ and $M1$ strength functions derived in the Sect.~\ref{sect:gsf},
assuming that the $M1$ spin-flip resonance is given either by the D1M+QRPA or the
Lorentzian strength and including an additional $M1$ upbend. In all cases,
the $\gamma$SF is firmly constrained by our dipole strength, with its lower and
upper value determined on the basis of the uncertaintites affecting not only
the $M1$-$E1$ decomposition, but also the average radiative width $\langle\Gamma_{\gamma 0}\rangle$
and s-wave resonance spacing $D_0$. Similarly, the NLD as derived in Sect.~\ref{sect:nld} are
modelled either by the HFB plus combinatorial model or the CT formula, and in each case
constrained on the experimental $D_0$ value with its upper and lower values. It should
be mentioned that the upper (lower) limit for the NLD ({\it i.e.} lower (upper) value of
the experimental $D_0$) is directly correlated to the upper (lower) limit for the
derived dipole $\gamma$SF, as constrained by the $\langle\Gamma_{\gamma 0}\rangle$.
The careful account of all these
uncertainties is translated into the hashed area displayed in  Fig.~\ref{fig:xs_Zr}.

A similar comparison is made for the Maxwellian-averaged cross sections in Fig.~\ref{fig:macs_Zr}.
The upper cross sections are found  with the upper value of the $\gamma$SF obtained with
the D1M+QRPA model of the $M1$ strength, while the lower cross section corresponds to the
lower value based on the $M1$ Lorentzian representation.

The main uncertainties in the present
analysis stem from the $E$1--$M$1 decomposition as well as the normalization of the experimental
$\gamma$SF. Note that the intrinsic model uncertainties, using all available NLD and $\gamma$SF
models in TALYS, yield a factor $\sim 10$ between the minimum and maximum $(n,\gamma)$ cross
sections in this mass region~\cite{larsen-yttrium}. Thus, although our indirect method gives
a rather large error band, it is still a significant improvement compared to the
range of possible values from the unconstrained model predictions.
This analysis shows that the NLD and $\gamma$SF derived in the present work are
fully compatible with the experimental radiative neutron capture cross sections
and can therefore be expected to be a good representation of the statistical properties
of the de-exciting compound nuclei $^{91}$Zr and $^{92}$Zr.

\section{Summary and outlook}
NLDs and $\gamma$SFs of $^{91,92}$Zr have been extracted from particle-$\gamma$ coincidence
data using the Oslo method. The data are normalized to neutron-resonance data and $(\gamma,n)$
cross section data, taking into account systematic errors due to uncertain nuclear spin distributions
as well as uncertainties in the extraction procedure and the external normalization data. 
Moreover, the $\gamma$SFs are decomposed into their $E1$ and $M1$ components based on state-of-the-art 
microscopic calculations of the $M1$ strength, as well as a phenomenological description of the $M1$ spin-flip 
resonance guided by previous $(p,p')$ measurements.

Our data, including all the possible normalization uncertainties, have been used as input for
calculating $^{90,91}$Zr$(n,\gamma)$ cross sections and MACS of relevance for the $A\sim90$ $s$-process peak.
We found that our indirect method of determining the MACS is fully compatible with 
direct measurements, giving confidence that this approach is capable of providing reasonable
cross sections for cases where direct measurements are not available, such as the branch-point nucleus $^{95}$Zr.
In the future, we will perform experiments at OCL to measure the $^{96}$Zr NLD and $\gamma$SF to 
deduce a first experimentally constrained $^{95}$Zr$(n,\gamma)$ cross section and MACS.

\acknowledgements

The authors wish to thank E.~A.~Olsen, A.~Semchenkov and J.~Wikne at the
Oslo Cyclotron Laboratory for providing excellent experimental conditions.
We are deeply grateful to A.~B\"{u}rger for his significant contribution to this work.
S.~S. gratefully acknowlegde funding by the Research Council of Norway (NFR),
project grant no. 210007.
A.~C.~L. gratefully acknowledges financial support from the Research Council of Norway,
project grant no. 205528 and from the ERC-STG-2014 under grant agreement no. 637686.
 S.~G. acknowledges the support from the F.R.S.-FNRS. A.~V. acknowledges the grant from Deparment of Energy no.~DE-NA0002905.

\end{document}